\documentclass[prl,floatfix,superscriptaddress,amsmath,amsfonts,twocolumn]{revtex4-1}
\usepackage{hyperref}
\pdfoutput=1
\usepackage{graphicx}
\usepackage{epsfig}

\usepackage{amssymb}
\usepackage{xcolor}

\newcommand{\PT}{{\cal PT}}

\newcommand{\tq}{\tilde{q}}

\DeclareMathOperator\sgn{sgn}

\usepackage[normalem]{ulem}
\hypersetup{backref,colorlinks=true,linkcolor=black,citecolor=black,urlcolor=dgreen,linktocpage=true,breaklinks=true}

\definecolor{dgreen}{rgb}{0.0, 0.0, .4}

\begin{document}

\title{\large Coherent-perfect-absorber and laser for bound states in a continuum}
\author{ Bikashkali Midya}
\affiliation{Institute of Science and Technology Austria, AM Campus 1, 3400 Klosterneuburg, Austria}
\author{ { Vladimir V. Konotop}}
\affiliation{Centro de F\'isica Te\'orica e Computacional and Departamento de F\'isica, Faculdade de Ci\^encias, Universidade de Lisboa, Campo Grande 2, Edif\'icio C8, Lisboa 1749-016, Portugal \vspace{.25cm}}

\begin{abstract}
 { It is shown that two fundamentally different phenomena, the bound states in continuum and the spectral singularity (or time-reversed spectral singularity), can occur simultaneously.  This can be achieved, in particular, in a rectangular core dielectric waveguide with embedded active (or absorbing) layer. In such a system a two-dimensional bound state in a continuum is created in the plane of a waveguide cross section and it is emitted or absorbed along the waveguide core. The idea can be used for experimental implementation of a laser or a coherent-perfect-absorber for photonic bound state that resides in a continuous spectrum.}
\end{abstract}
\maketitle

Unusual points in spectra of optical systems, i.e. isolated points, which do not exist in a generic situation and may require special engineering of the system, attract increasing attention during the last years. For either Hermitian or non-Hermitian systems these are, in particular, the bound states in continuum (BIC), which were predicted almost a century ago~\cite{Neumann1929} and acquired great experimental and practical importance~\cite{opt1,opt2,Rivera2016,opt3,opt4} in recent years (see also~\cite{BIC_review} and reference therein). In the case of non-Hermitian optical systems, the continuum spectrum can  {also} host either spectral singularities~\cite{Mostafazadeh2009,Longhi2010}, at which the system behaves like a laser, and time-reversed spectral singularities at which the system becomes a coherent-perfect-absorber (CPA) for light at a given wavelength~\cite{Chong2010,StoneScience,Baranov2017}.

From the mathematical point of view, either BIC or spectral singularities (for the sake of brevity under spectral singularities here we also understand time-reversed spectral singularities) represent isolated points   {inside the continuous} spectrum, requiring fine tuning of the parameters to be achievable in principle. From the physical point of view in the case of CPA and BIC the parameters of the scattering potential must be chosen to ensure a delicate condition of destructive interference of outgoing radiation completely canceling it at the infinities. Thus, it looks very unlikely to satisfy the conditions for a BIC and either laser or CPA at the same wavelength (just because this would require too many free system parameters). The aim of this Letter is to show that in practice this is not necessarily so, and the conditions for a BIC and for a spectral singularity can be met simultaneously. In other words one can create a laser or a CPA for a BIC.  We explore the waveguide geometry where a two-dimensional (2D) BIC is created in the plane orthogonal to the waveguide axis; {and} a spectral singularity  {is} found in the spectrum of the longitudinal waves. Thus although both phenomena occur   simultaneously, they remain conceptually different. Nevertheless, as shown below, the phenomena of BIC and spectral singularity are correlated for a guided wave due to the energy exchange between the transverse and longitudinal directions in a waveguide.
  
Of late, a BIC laser was  demonstrated {experimentally}~\cite{Kante2017}, although using a mechanism which differs from the one presented in this Letter. A two-dimensional lattice was illuminated from one side,  {and} displayed lasing action {of} a BIC from the opposite side. The emission of light, which is considered here, is also orthogonal to the plane of the BIC localization, but it occurs in two opposite directions simultaneously and relies on simultaneous existence of a BIC and a spectral singularity in the same setting.  Also, recently, it has been found numerically~\cite{Jianke} and explained analytically to be a  quite general phenomenon~\cite{KZ}, that localized bound states having real part of the propagation constant in the continuous spectrum, can emerge if the continuous spectrum has a self-dual singularity. Such bound states however being similar to BICs, are not exact BICs in their canonical definition: due to presence of imaginary components of the propagation constant such modes either grow or decay, remaining localized. The beams emitted or absorbed by the spectral singularity, as reported in this Letter, are the authentic BICs, having constant amplitude and residing in the continuous spectra.

\begin{figure}[h!]
\centering
\includegraphics[width=.73\columnwidth]{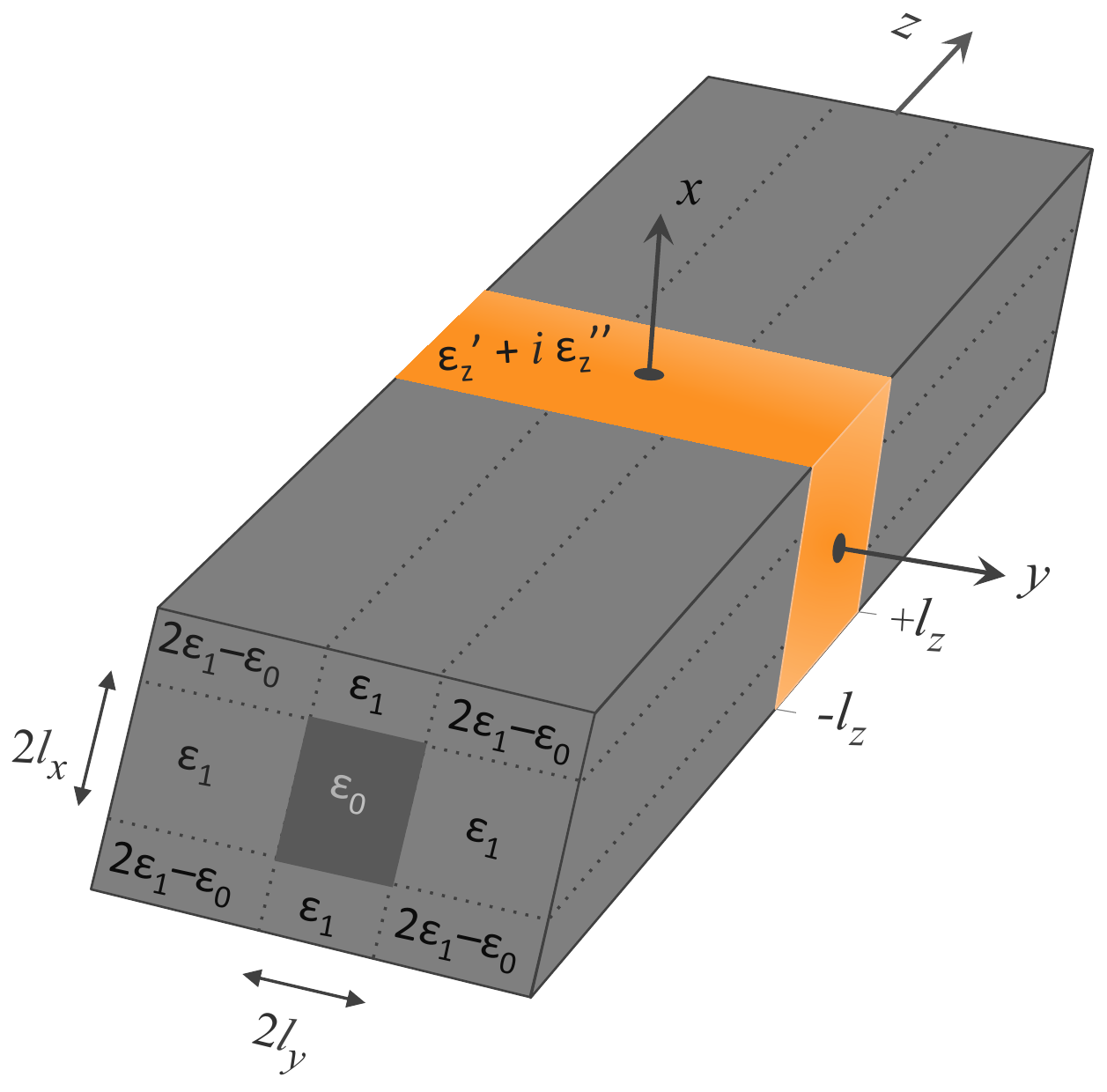} 
\caption{A schematic representation of the waveguide which is considered in the text. The distribution of the dielectric constants in different parts of the waveguides, as well as their dimensions, are indicated. } \label{fig-diagram}
\end{figure}

Consider a three-dimensional (3D) rectangular waveguide with a core which is infinitely extended in the $z$-direction and finite in the transverse $(x,y)$ plain, as Fig.~\ref{fig-diagram} shows. Assume that the waveguide is characterized by the dielectric constant $\varepsilon(\vec{r}) = \varepsilon_x(x) + \varepsilon_y(y) + \varepsilon_z(z)$ such that 
\begin{align} \label{Eq-pot}
\varepsilon_\alpha(\alpha) = \left\{\begin{array}{c} \varepsilon_0/2, \quad\quad~~~ |\alpha|< \ell_\alpha\\ \varepsilon_1-\varepsilon_0/2, \quad |\alpha| > \ell_\alpha\end{array}\right.
\end{align}
where $\alpha = x, y$, $\varepsilon_0 > \varepsilon_1$, $\ell_{x,y}$ are the waveguide dimensions along the $x$ and $y$ directions. The dielectric constant, $\varepsilon_z$, in the longitudinal direction, is a complex number $\varepsilon_z=\varepsilon_z'+i\varepsilon_z''$ inside the scattering region situated in $|z| < \ell_z$, and zero outside this domain. The imaginary part represents either absorption ($\varepsilon_z''>0$) or gain ($\varepsilon_z''<0$) of the scattering layer.

Thus, outside the layer,  $|z|>\ell_z$, the total dielectric constant of the rectangular core (i.e. at $|x| < \ell_x$ and $|y| <\ell_y$) is given by $\varepsilon_0$; whereas for the cladding it is $\varepsilon_1$, and $2\varepsilon_1 - \varepsilon_0$ at the four corners (see fig.~\ref{fig-diagram} for the schematic distribution of the total dielectric constant in the $xy$-plane). Such type of the waveguide configuration is well studied since long ago~\cite{Marcatili1969,Kumar1983}.

In such a setting, a BIC CPA or a BIC laser is constructed in a two-step procedure. First, a BIC is constructed that is in a normalizable transverse mode inside the radiation spectrum. Second, such a BIC mode is allowed to interact with  the (absorbing or active) layer located in the interval $|z| \le \ell_z$. Let $M$ be the $2\times 2$ transfer matrix of the layer ({effective} one-dimensional scattering along the $z-$axis is considered). Then, the CPA and laser  correspond to real zeros of the transfer matrix elements $M_{11}$ and $M_{22}$.  In the following we describe these two steps for {realistic} examples.

To introduce discrete and BIC guided modes {we note that} outside the scatterer (where $\varepsilon_z = 0$), the dielectric permittivity is constant in each of the layers. Therefore, the spatial distribution of the magnetic field, $\vec{H}(\vec{r},t) = \vec{H}(r) e^{-i \omega t}$, of a monochromatic wave having the frequency $\omega$, solves the Helmholtz equation $\nabla^2 \vec{H} + k^2 \varepsilon \vec{H} =0,$ where $k = \omega/c$ is the wavenumber in vacuum. Considering a mode propagating along $z$ direction, i.e. $\vec{H}(\vec{r}) = e^{i q z} \left(\psi, 0, \frac{i}{q} \frac{\partial \psi}{\partial x}\right)$, with $q$ being the $z-$component of the wavevector, we obtain that  $\psi(x,y)$ satisfies 
\begin{align}\label{Eq-1}
\frac{\partial^2 \psi}{\partial x^2} + \frac{\partial^2 \psi}{\partial y^2} + \left( k^2 \varepsilon - q^2\right)\psi = 0.
\end{align} 
 It is worth mentioning here that this type of modes are neither TE nor TM, but are hybrid modes known as $E^y_{mn}$ (as the only dominating field components are $E_y$ and $H_x$)~\cite{Marcatili1969}. Nevertheless, in the limit of negligible dielectric permittivity difference: $|\varepsilon_0 - \varepsilon_1|\ll \varepsilon_{0,1}$, the electric field becomes a plane wave linearly polarized along $y-$ direction.

The guided modes, are given by solutions of Eq.~(\ref{Eq-1}) in the form  $\psi_{mn}(x,y) = X_m(x) Y_n(y)$, with $m$ and $n$ integers. The functions $X(x)$ and $Y(y)$  are obtained by solving $d^2X/dx^2 + (k^2 \varepsilon_x - q_x^2) X =0$ and $d^2Y/dy^2 + (k^2 \varepsilon_y - q_y^2) Y =0$ with the auxiliary spectral parameters $q_{x,y}$ satisfying $q_{x}^2 + q_{y}^2 = q^2$. The modes in both directions can be either symmetric,  
\begin{eqnarray} \label{Eq-sol1}
X(x) = \left\{\begin{array}{ll} \cos (k_x x), & |x| < \ell_x \\ 
\cos (k_x \ell_x)  e^{- \kappa_x (|x| - \ell_x)},  & |x| > \ell_x\end{array}\right.
\end{eqnarray}
or antisymmetric,  
\begin{eqnarray} \label{Eq-sol2}
X(x) = \left\{\begin{array}{ll} \sin (k_x x), & |x| < \ell_x \\ 
\sgn(x) \sin (k_x \ell_x) ~ e^{- \kappa_x (|x| - \ell_x)}, & |x| > \ell_x \end{array}\right.
\end{eqnarray}
in the $x$-direction. Similar expressions are valid for the $y-$direction with replacements $X\leftrightarrow Y$ and  $x\leftrightarrow y$. Here $k_\alpha = \sqrt{k^2 \varepsilon_0/2 -q_\alpha^2}$  and $\kappa_\alpha = \sqrt{k^2(\varepsilon_0/2 -\varepsilon_1) + q_\alpha^2}$,  with $\alpha=x,y$, are subject to the constraint $k_x^2 + k_y^2 =k^2$ with $k^2 (\varepsilon_0-1) = q^2$, and no normalization is used.

\begin{figure}[t!]
\centering
\includegraphics[width=.9\columnwidth]{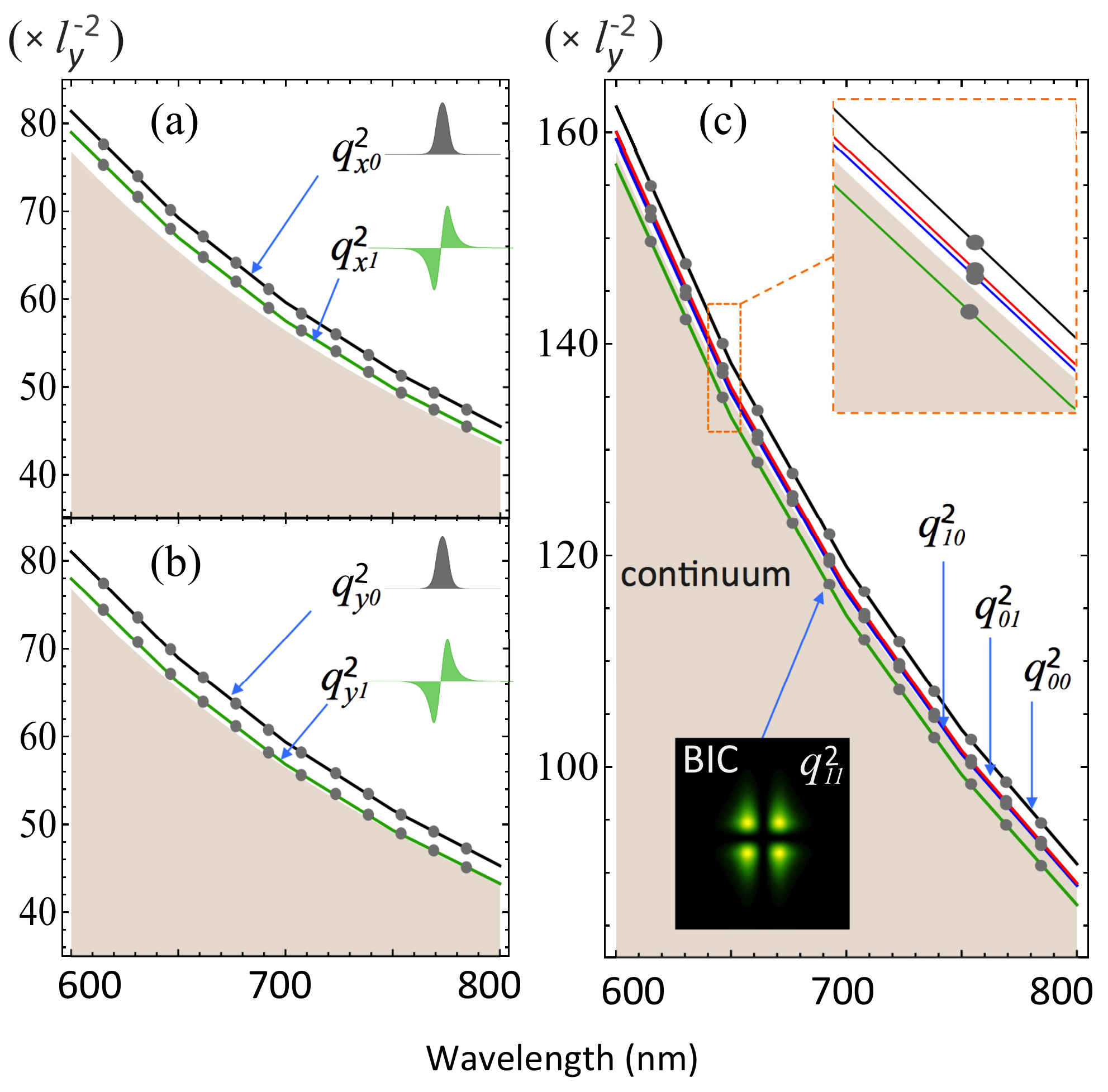}
\caption{In (a) and (b) we show the  eigenvalue spectrum, together with mode profiles, of the separable waveguide along $x$ and $y$ directions respectively; whereas in (c) we show the propagation constant along the longitudinal direction. Note that all the $q$'s are in the unit of $[\ell_y^{-1}]$. The mode $q_{11}^2$ lies inside  the continuum for the whole range of wavelength as shown in (c). Therefore this mode is the BIC. The corresponding real part of the magnetic field is shown in the inset (at the bottom) of (c), when $\lambda = 700~$nm was considered. In the inset (at the top) of (c) we  show the zoom of the spectrum near $\lambda = 650~$nm. We considered $\ell_y = \ell_x/1.25 = 1~\mu$m, $\varepsilon_0 = 1.5$, and $\varepsilon_1 = 1.45$ for all the plots.} \label{fig-1}
\end{figure}

Since the continuity of the $x$-component of the magnetic field is assured by the mode expressions (\ref{Eq-sol1}) or (\ref{Eq-sol2}), the dispersion relations for the spectral parameters $q_{x,y}$ are obtained from the continuity of  $z$-components of the magnetic and electric field at the core-cladding interface. They read
\begin{align}\label{Eq-xdispersion}
k_x  \tan (\ell_x k_x) = \kappa_x, \qquad \varepsilon_1 k_y  \tan (\ell_y k_y) = \varepsilon_0 \kappa_y,
\end{align}
for the symmetric modes and    
\begin{align}\label{Eq-ydispersion}
k_x  \cot (\ell_x k_x)  = - \kappa_x,  \quad\quad  \varepsilon_1 k_y  \cot (\ell_y k_y)  = - \varepsilon_0 \kappa_y,
\end{align}
for the antisymmetric modes.  
Note that for practical applications it is desirable to consider same dielectric constant for the whole cladding, i.e., the four corners of the cladding also having the dielectric constant as $\varepsilon_1$ (see Fig.~\ref{fig-1}(a)). Then, using  the smallness of the perturbation term $\varepsilon_0 -\varepsilon_1$ (in realistic settings $\varepsilon_0 \approx \varepsilon_1$), one can construct the solutions at the corners, from those of $\psi_{mn}$, by using  $(\varepsilon_0 -\varepsilon_1)$ as a perturbation on top of the potential defined in Eq.~(\ref{Eq-pot})~\cite{Kumar1983}. 

 The symmetric and antisymmetric transverse modes in each of the directions are further classified into:   the discrete set of guided modes with $k^2(\varepsilon_1-\varepsilon_0/2)<q_\alpha^2<k^2\varepsilon_0/2$, which are confined to the waveguide core; and  evanescent modes which propagate in cladding for $q_\alpha^2 < k^2 (\varepsilon_1- \varepsilon_0/2)$ representing the continuous spectrum.  The boundary between discrete and continuum spectra is given by  $q_{\rm b}^2=k^2 (\varepsilon_1- \varepsilon_0/2)$ (in our case it is same for both $x$ and $y$ directions).

Let us denote the discrete modes as $q_{\alpha m}$ and order them as follows:  $k^2 \varepsilon_0/2 > q_{\alpha 0}^2 >q_{\alpha 1}^2 \cdots >q_{\alpha M_\alpha}^2 > q_{\rm b}^2$, where $M_{\alpha}+1$ are the numbers of the guided modes confined in $\alpha=x,y$ direction. The numbers $M_x$ and $M_y$ may be different in view of different $\ell_x$ and $\ell_y$.   Now, note that a hybrid mode $mn$ becomes evanescent if either of $X_m$ or $Y_n$  or both becomes evanescent. Then the continuum for the guided hybrid modes starts at the critical value of the propagation constant
\begin{align}
q^2 = q_{\rm c}^2 = {max}\{q_{x0}^2 + q_{\rm b}^2,~q_{y0}^2 + q_{\rm b}^2\}.
\end{align}
The BICs (if they exist) for such a waveguide correspond to pairs  $(m,n)$ such that the corresponding propagation constant $q^2_{mn} = q_{xm}^2 + q_{yn}^2$ lies below  the continuum threshold ${q_{\rm c}^2}$. This situation is explained in the example below and in the figure \ref{fig-1}(c). The construction of this type of ``separable" BIC was theoretically predicted earlier in quantum mechanics~\cite{Robnik1986}. However, the condition for a BIC to exist, and the BIC threshold differ significantly in the optical settings as explained above.

Now, consider a wave with a wavenumber $k$, corresponding to a BIC mode with $m=m_0$ and $n=n_0$ i.e. with  {$q_{\!_{BIC}}=q_{m_0n_0} < q_{\rm c}$}, which is propagating in the waveguide. We are interested in the stationary scattering of this BIC by an active or dissipative layer placed in $|z| <\ell_z$. The right (``$+$") and left (``$-$") propagating modes {in this BIC channel}, can be  searched in the from 
$ {\vec{H}^\pm}=\vec{{\mathcal{H}}}^\pm e^{\pm iq_{\!_{BIC}}z}$, where
$$\vec{{\mathcal{H}}}^\pm  =  \left(\psi_{\!_{BIC}}, 0, \pm \frac{i}{q_{\!_{BIC}}} \frac{\partial \psi_{\!_{BIC}}}{\partial x}\right).$$ 
The field inside the scattering layer is {given by} the superposition of the components:
$$\vec{\tilde{\mathcal{H}}}^\pm  =  \left(\psi_{\!_{BIC}}, 0, \pm \frac{i}{\tq_{\!_{BIC}}} \frac{\partial \psi_{\!_{BIC}}}{\partial x}\right),\quad \tq_{\!_{BIC}} =\sqrt{q_{\!_{BIC}}^2+ k^2\varepsilon_z}.$$
Then the full scattering solution can be written in the form 
\begin{equation}\label{scat-state}
\vec{H}(\vec{r}) =\left\{\begin{array}{ll} a   e^{i q_{\!_{BIC}} (z+\ell_z)}~\vec{\mathcal{H}}^+ + b e^{-i q_{\!_{BIC}} (z+\ell_z)}~\vec{\mathcal{H}}^-, & z < -\ell_z\\ 
g e^{i  \tq_{\!_{BIC}} z} \vec{\tilde{\mathcal{H}}}^+ + h e^{-i \tq_{\!_{BIC}}  z} \vec{\tilde{\mathcal{H}}}^- , & |z| < \ell_z \\
c e^{i q_{\!_{BIC}} (z-\ell_z)}~\vec{\mathcal{H}}^+ + d e^{-i q_{\!_{BIC}} (z-\ell_z)}~\vec{\mathcal{H}}^- , & z > +\ell_z\end{array}\right.
\end{equation}  

\begin{figure}[t!]
\centering
	\includegraphics[width=.40\textwidth]{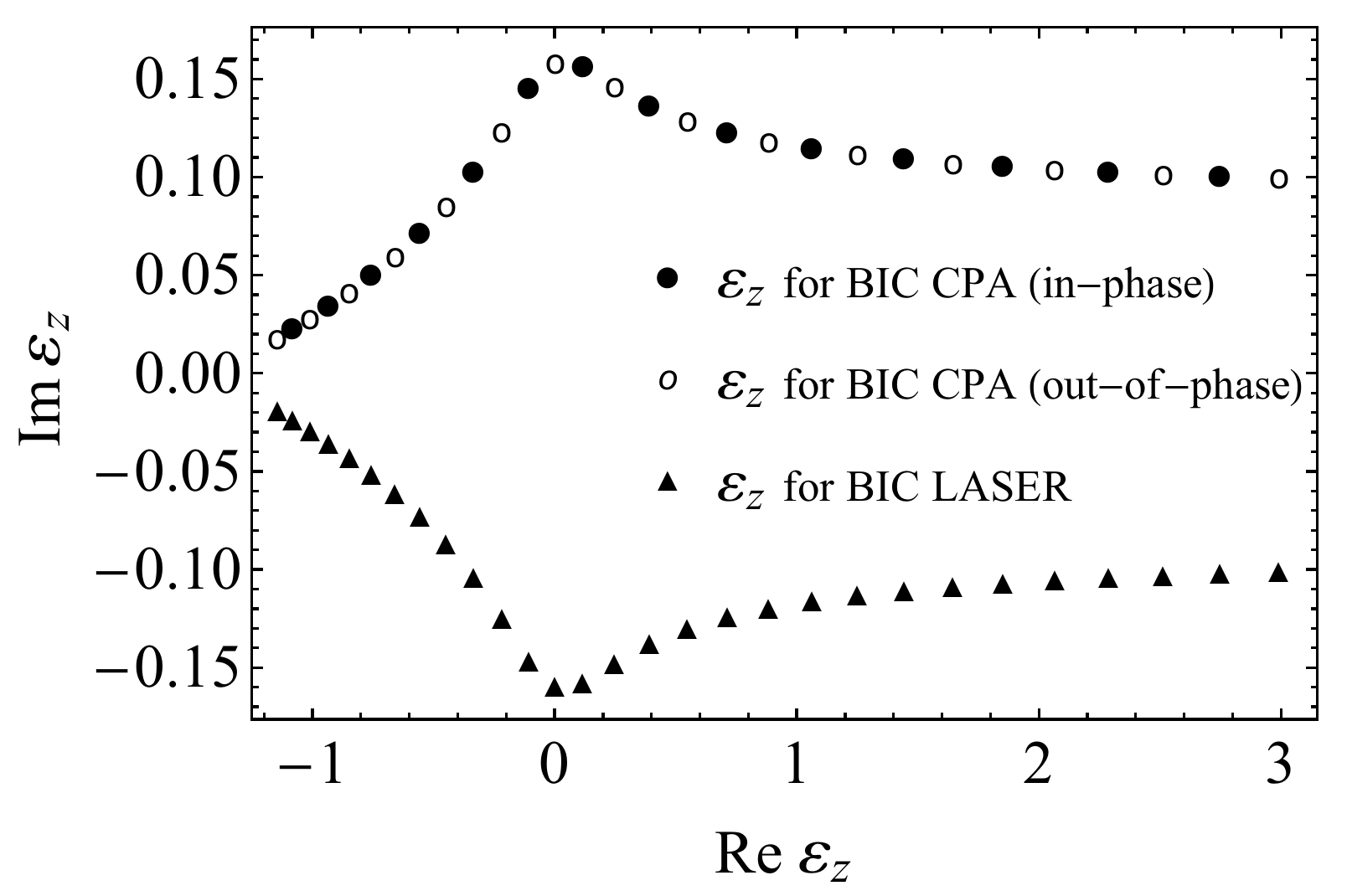}
	\caption{The distribution of $\varepsilon_z$ in the complex plane has been shown for which BIC CPA and BIC laser transition occur when $\ell_z = 3\mu$m, $\lambda=700$nm, $\varepsilon_0 = 1.5$, and $\varepsilon_1 = 1.45$. The solid circle and empty circle emphasis the BIC CPA  for in-phase and out-of-phase inputs (the phase relations are necessary for observation of the phenomenon) of equal intensities, respectively.} \label{fig-3}
\end{figure}

Due to the orthogonality of the transverse modes, the BIC will not excite other modes inside the layer except the incident and outgoing ones (both of which are the transverse mode $m_0,n_0$ in our case). Now the  transfer matrix $M$ is defined through the relation $(c,d)^T=M\,(a,b)^T$ (here $T$ stands for transpose), and its components are computed as
\begin{align}
M_{11} =  \cos (2 \ell_z \tq_{\!_{BIC}}) + i ~\frac{q_{\!_{BIC}}^2 + \tq_{\!_{BIC}}^2}{2 q_{\!_{BIC}} \tq_{\!_{BIC}}} \sin (2 \ell_z\tq_{\!_{BIC}}) \label{Eq-M11}\\
M_{12} = - M_{21} =i \frac{q_{\!_{BIC}}^2 - \tq_{\!_{BIC}}^2}{2 q_{\!_{BIC}} \tq_{\!_{BIC}}} ~\sin (2 \ell_z \tq_{\!_{BIC}})\\
M_{22} =  \cos (2 \ell_z \tq_{\!_{BIC}}) - i ~\frac{q_{\!_{BIC}}^2 + \tq_{\!_{BIC}}^2}{2 q_{\!_{BIC}} \tq_{\!_{BIC}}} \sin (2 \ell_z \tq_{\!_{BIC}}) \label{Eq-M22}
\end{align}
 {The left (``$L$") and right (``$R$") reflection and transmission amplitudes are related to the transfer matrix elements by: $t_L=t_R=t=1/M_{22}$, and $r_L=-M_{21}/M_{22}$, and $r_R=M_{12}/M_{22}$; in our case $r_L=r_R=r$. 
 	
Two remarks are in order. First, a CPA is characterized  by modes propagating  {\em only} towards the layer (in our case $b=c=0$). A laser is characterized by modes  propagating {\em only} outwards the layer (in our case $a=d=0$). Since in both cases these are oppositely propagating waves, we conclude that either  CPA or laser are realized by two modes with different polarizations. Second, since the link between the scattering data and elements of the transfer matrix is standard, using the conventional arguments~\cite{Chong2010} one conclude that either for a CPA or for a laser, $M_{12}^2=1$, and the two beam involved (absorbed or emitted) have equal amplitudes and they must be either in-phase ($0$ phase difference) or out-of-phase ($\pi$-phase difference).

\begin{figure}[t!]
\centering
	\includegraphics[width=.5\textwidth]{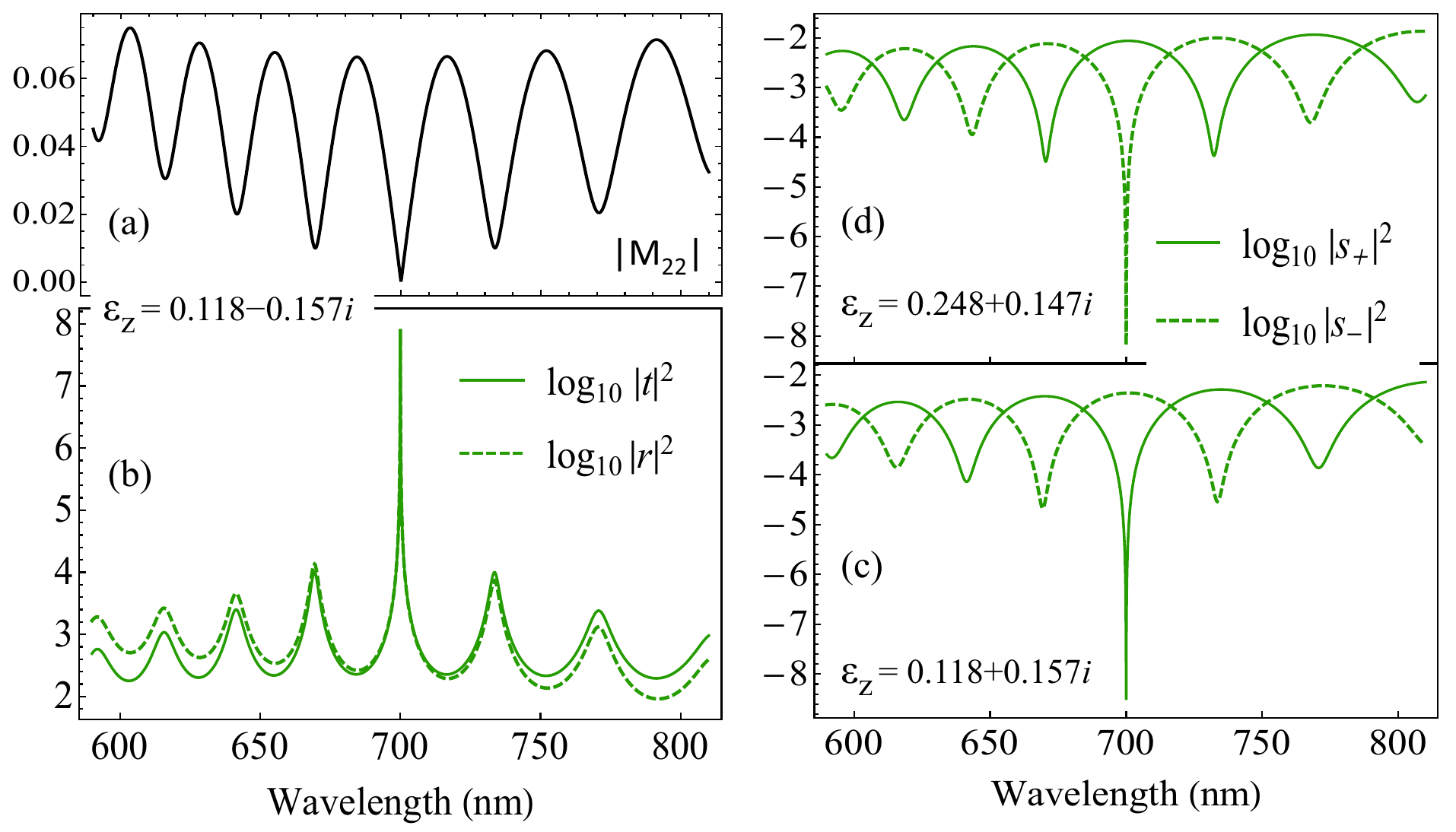}
	\caption{(a) The transfer matrix element $M_{22}$ is shown as a function of wavelength $\lambda = 2\pi/k$. (b) The transmission and reflection coefficients are shown (in semi-log plot) when BIC laser action occurs for $\varepsilon_z =  0.118-0.157i$. The BIC CPA normalised output intensities,$|s_\pm|^2$, for in-phase and out-of-phase input beams are shown (in semi-log plot) for $\epsilon_z=0.118+0.157i$ in (c), and for $\epsilon_z=0.248+0.147i$ in (d). In all these plots we considered the parameter $\ell_z = 3\ell_y$.}\label{fig-4}
\end{figure}

Now we turn to concrete examples illustrating the described idea. For the sake of concreteness we consider $\varepsilon_0 = 1.5$ and $\varepsilon_1 = 1.45$. The rectangular core dimension is fixed to $\ell_y=\ell_x/ 1.25=1\mu$m. For these values of the parameter we  have obtained the spectrum for the waveguide for a wide range of wavelength. Corresponding eigenvalues and mode profiles are shown in Fig.~\ref{fig-1}.  The bound mode $q_{11}^2$ lies inside the continuum and thus corresponds to a BIC. It is evident from the figure that this BIC is robust against the frequency tuning. In particular for $\lambda=700$nm, we have  $q_{x0}^2=59.635$, $q_{x1}^2 =57.526$, $q_{y0}^2=59.343$, and $q_{y1}^2 =56.802$ (all measured in the units of~$\ell_y^{-2}$). The continuum for radiation modes starts at $q_c^2 = 116.03~\ell_y^{-2}$. The mode $q_{11}^2 = 114.329\ell_y^{-2}$ lies inside the continuum, and thus is a BIC.

 When the gain (or loss) layer is illuminated by such a BIC mode, there exist certain values of $\varepsilon_z$ for which one can obtain the laser condition $M_{22} = 0$ at the $700$nm wavelength. For $\ell_z=3\ell_y$ (this particular value is chosen for the sake of concreteness only, other choice of $\ell_z$ is also possible), 
 we numerically obtain several values of $\varepsilon_z$ for which laser action occurs, and are shown in Fig.~\ref{fig-3} with triangles.  Similarly, by solving $M_{11} = 0$, we obtain $\varepsilon_z$'s for a CPA (at $\lambda=700$nm) which are shown in Fig.~\ref{fig-3} by solid and empty circles for in-phase and out-of-phase modes, respectively. Note that the distribution for $\epsilon_z$, for CPA and laser, is mirror symmetric with respect to the real axis. The strength of the gain or loss necessary for lasing or perfect absorption decreases with the increase of the modulus of the real part of $\varepsilon_z$ (recall that this value is not the real part of the total dielectric constant but rather a deviation from the dielectric constant of the layer region, which is $\varepsilon_0+\varepsilon_z$, from that of the waveguide kernel).

In Fig.~\ref{fig-4}, we show an example of the dependencies of the transfer matrix element $M_{22}$ [panel (a)], and of the scattering coefficients $|t|^2$ and $|r|^2$ [panel (b)], on the wavelength near the laser threshold. In the panels (c) and (d) of  Fig.~\ref{fig-4}, we show the normalized BIC output intensities, which are defined by $|s_\pm|^2$ with $s_\pm= r\pm t$ being the eigenvalues of the respective scattering matrix of the in-phase and out-of-phase eigenmodes for a CPA~\cite{Chong2010}. In Fig.~\ref{fig-4}(c) the results are shown when CPA occurs for in-phase beams [at  $\epsilon_z=0.118+0.157i$], and ~\ref{fig-4}(d) corresponds to CPA for  out-of-phase beams [at $\epsilon_z=0.248+0.147i$]. While the resonant curves, shown in panels (b), (c) and (d), look quite commonly, they have a peculiarity in our case.  The mode remains a BIC in spite of tuning the wavelength. This is possible due to the change of the mode profile ($q_{11}$ mode in our case) with change of $k$. In other words, a BIC absorbed or emitted by the CPA or laser (the zero width resonance) has different spatial profile, and as a consequence different polarization, as compared with all BICs corresponding to other wavelengths.

To conclude, the reported results are twofold. First, we have shown how to construct a BIC in a rectangular core of a two-port dielectric waveguide system. Second, we described how to realize a BIC CPA and a BIC laser, by embedding either absorbing or active layer of a finite length along the propagation direction. The mechanism is explained with experimentally feasible examples. Note that the BIC created inside the cavity is actually 2D in nature, localized in the plane transverse to the propagation direction. Thus, the corresponding laser, although 3D, is a composition of the 2D BIC and the plane wave (along the longitudinal direction). In this respect, the corresponding laser/CPA is a 2D BIC laser/CPA. As a final remark, the presented waveguide geometry potentially satisfy the condition of self-dual spectral singularity~\cite{Mostafazadeh2012} at which $\PT$-symmetric laser-absorber~\cite{Longhi2010,Liang2016} is possible, although requires further investigation.


{ \it  B. Midya acknowledges financial support by the People Programme (Marie Curie Actions) of the European Union's Seventh Framework Programme (FP7/2007-2013) under REA grant agreement No. [291734].}


\end{document}